\documentclass[pra,aps,twocolumn,showpacs,groupauthors,nofootinbib]{revtex4}

\usepackage{amssymb}
\usepackage[english]{babel}
\usepackage[ansinew]{inputenc}
\usepackage{amsfonts}
\usepackage{amsmath}
\usepackage{graphicx}
\usepackage{bm}
\usepackage{color}
\usepackage{rotating}

\newcommand{\ket}[1]{|{#1}\rangle}
\newcommand{\bra}[1]{\langle{#1}|}

\newcommand{\ave}[1]{\langle{#1}\rangle}

\begin{document}


\title{Estimation of classical parameters via continuous probing of complementary quantum observables}


\author{Antonio~Negretti$^{1,2}$ and Klaus~M\o lmer$^3$}
\affiliation{$^1$Institut f\"ur Quanteninformationsverarbeitung, Universit\"at Ulm,
Albert-Einstein-Allee 11, D-89069 Ulm, Germany\\
$^2$Zentrum f\"ur Optische Quantentechnologien, Universit\"at Hamburg, The Hamburg Centre for Ultrafast Imaging, Luruper Chaussee 149, D-22761 Hamburg\\
$^3$Lundbeck Foundation Theoretical Center for Quantum System Research\\
Department of Physics and Astronomy, University of Aarhus\\
DK-8000 Aarhus C, Denmark}
\date{\today}


\begin{abstract}
We discuss how continuous probing of a quantum system allows estimation of unknown classical parameters embodied in the Hamiltonian of the  system. 
We generalize the stochastic master equation associated with continuous observation processes to a Bayesian  filter equation for the  probability 
distribution of the desired parameters, and we illustrate its application by estimating the direction of a magnetic field. In our example, the 
field causes a ground state spin precession in a two-level atom which is detected by the polarization rotation of off-resonant optical probes, 
interacting with the atomic spin components.
\end{abstract}




\maketitle


\section{Introduction}
\label{sec:intro}

High precision metrology with quantum systems is a research field of high current activity. Through the quantization of their energy levels, elementary quantum systems provide fundamental time and frequency standards and, due to the highly developed means for preparation, control, and detection of these systems, they serve as excellent probes of various perturbations such as applied electric and magnetic fields, and inertial effects associated with rotation, acceleration, and classical or relativistic gravitational effects.

The theoretical research proceeds along different directions according to the different measurement schemes. Thus, for experiments where a quantum system is subject to a perturbation for a given short duration of time, the search for the initial quantum states on which different values of the perturbation leads to the most distinguishable outcomes has promoted the use of concepts such as the Fisher information and the Cramer-Rao bound~\cite{Braunstein1994}, and has identified squeezed states,
Schr\"odinger cat-like states and generalizations hereof as useful resource states in metrology. Along a different path, measurements that occur continuously in time, such as continuous wave laser spectroscopy, are made subject to analyses, that serve to exhaust the information about the desired parameters from the entire sequence of measurement data. While a simple relationship between the average signal, e.g., an absorption profile, and an unknown physical parameter provides a relatively straightforward method for estimation, the systematic extraction of
a reliable error-bar on the estimate is more challenging~\cite{Molmer2004,Maybeck1982,Tsang2012}.

In this work we present such an analysis for the non-trivial case of the continuous probing of a single quantum system. We detail a Bayesian analysis, which treats  our description of unknown parameter values and the unknown state of the quantum system on an equal footing and where any data received serves to update our prior estimate of the parameters of interest. This idea has been previously applied to the case of quantum-non-demolition (QND) probing of quantum systems~\cite{Geremia2003,Molmer2004}, where it is largely equivalent to a Kalman filter~\cite{Maybeck1982}.  The formalism presented here, however, is more general, and new physical features appear, when not only a single QND variable is being detected.

The paper is organized as follows: In Sec. II we present our general quantum filter equation and we give an explicit derivation for our specific physical model system. In Sec. III we introduce the unknown classical parameters and we show how their representation as effective incoherent quantum degrees of freedom
augments the quantum filtering equation to automatically provide a Bayesian update formula for the unknown classical parameters. In Sec. IV we present numerical simulations and we show how the fluctuating measurement signals of optical probes interacting with an atomic spin gradually filters the probability distribution of a magnetic field with known strength, but unknown direction, and how it ultimately reveals this direction with good precision. In Sec. V we conclude and discuss the outlook and perspectives of our work. Details about the derivation of the augmented stochastic master equation (SME) and on its numerical solution are provided in the appendix.


\section{Optical probing of a single atomic spin}
\label{sec:four}

\subsection{General quantum filtering equation}

A quantum system with Hamiltonian $\hat{H}$ subject to Markovian damping, described by Lindblad operators
$\hat{\cal O}_j$ and rates $\Gamma_j$, is described by a reduced system density matrix $\hat\varrho^s$, which obeys the master equation

\begin{align}
\label{eq:mestandard}
\mathrm d \hat\varrho^s &= \left(\frac{i}{\hbar}
[\hat\varrho^s,\hat H]
+ \sum_j\Gamma_j \mathcal{D}[\hat{\cal O}_j]\hat\varrho^s
\right)\mathrm d t,
\end{align}
where the operator $\mathcal{D}$ is defined as~\cite{Wiseman2010}

\begin{align}
\label{eq:D-def}
\mathcal{D}[\hat f]\hat\varrho^s_{\mathrm s} &= \hat f \hat\varrho^s_{\mathrm s} \hat f^{\dag}
- \frac{\hat f^{\dag}\hat f \hat\varrho^s_{\mathrm s} + \hat\varrho^s_{\mathrm s}\hat f^{\dag}\hat f}{2}.
\end{align}

Interaction with continuous quantized probe fields cause entanglement of the system with the fields which, if the field degrees of freedom are subsequently traced
out, is described by inclusion of further master equation terms $\mathcal{D}[\hat{\cal M}_n]\hat\varrho$ with Lindblad operators $\hat{\cal M}_n$ and effective interaction parameters $M_n$. The entanglement of the system and the fields, however, implies that measurements of the probe field variables after the interaction lead to a back-action on the state of the quantum system.

Subject to the quantum back-action due to continuous amplitude measurements on the probe field, the reduced density matrix of the quantum system obeys the following stochastic master equation

\begin{align}
\label{eq:sme1}
\mathrm d \hat\varrho^s &= \left(\frac{i}{\hbar}
[\hat\varrho^s,\hat H]
+ \sum_j\Gamma_j \mathcal{D}[\hat{\cal O}_j]\hat\varrho^s
+ \sum_{n} M_n \mathcal{D}[\hat{\cal M}_n]\hat\varrho\right)\mathrm d t
\nonumber\\
\phantom{=}&\!\!\!\!\!\!\!\!
+ \!\!\sum_{n} \!\sqrt{\eta_n M_n}\mathcal{H}[\hat{\cal M}_n]\hat\varrho^s \mathrm{d}W_n(t),
\end{align}
where the operator $\mathcal{H}$ is defined as~\cite{Wiseman2010}

\begin{align}
\label{eq:H-def}
\mathcal{H}[\hat f]\hat\varrho^{\mathrm s} &= \hat f \hat\varrho^{\mathrm s} + \hat\varrho^{\mathrm s} \hat f^{\dag}
- \ave{\hat f + \hat f^{\dag}}\hat\varrho^{\mathrm s},
\end{align}
and where d$W_n$ are infinitesimal Wiener processes accounting for the noisy contribution to the field amplitude measurement
\begin{align}
\label{eq:noisy-signal}
\mathrm{d}Y_n^D(t)=\eta_n\sqrt{M_n}\ave{\hat{\cal{M}}_n+\hat{\cal{M}}_n^{\dagger}}\mathrm{d}t + \mathrm{d}W_n(t).
\end{align}
In the equations (\ref{eq:sme1}) and (\ref{eq:noisy-signal}) the parameters $\eta_n$ account for the detector efficiencies and transmission losses of the probe beams between the system and the detector. If $\eta_n=0$, the corresponding probe field merely contributes extra damping and decoherence to the system, while if $\eta_n=1$, and in the absence of other damping or ineffective probing terms, the system may be described by a stochastic wave function rather than a density matrix~\cite{Carmichael1993,Dalibard1992,Molmer1993,Wiseman1993}.
The random Wiener noise increments d$W_n$ constitute the so-called innovation processes, i.e., they are the difference between the experimentally observed signals d$Y_n^D(t)$ and their quantum-mechanical expectation values $\eta_n\sqrt{M_n}\ave{\hat{\cal{M}}_n+\hat{\cal{M}}_n^{\dagger}}\mathrm{d}t$ determined from the current quantum state of the system. This difference amounts to the shot-noise in field amplitude measurements by homodyne detection, and the formalism can also be adapted to treat, e.g., photon counting measurements, where the innovation process is described by an (infinitesimal) Poisson process. In either case, the measurement back-action has only infinitesimal influence in every small time step while for certain measurements accumulated over time it ultimately causes the collapse on a random eigenstate normally attributed to the von Neuman projection postulate.

We have presented the quantum filtering equation in a general form following a Markovian and perturbative treatment of the interaction of the system with its environment. For every concrete physical example one has to validate such a treatment and to explicitly analyze the appropriate interactions and field measurement schemes to obtain the operators and parameters in Eq.~(\ref{eq:sme1}). In the next subsection we will recall such a derivation for the explicit case of an atom with a degenerate ground state that interacts with off-resonant probe laser fields.

\subsection{Dynamics of a two-state atom and an off-resonant laser field}
\label{sec:fourA}

The detection of optical phase shifts and polarization rotation is at the heart of spin squeezing and quantum entanglement schemes involving the collective spin degrees of freedom associated with large atomic ensembles~\cite{Julsgaard2001,Madsen2004,Kuzmich2004,Sherson2006}, and it also constitutes the basis for atomic magnetometers which today explore the quantum limits of resolution~\cite{Wasilewski2010,Shah2010,Koschorreck2010}. The spins in large atomic ensembles are well approximated as harmonic oscillator degrees of freedom and Gaussian approximations and classical filtering theory apply~\cite{Madsen2004,Stockton2004,Petersen2005}, while single atoms must be described by the full density matrix formalism, that we will address in the following.

We consider an atomic quantum system with degenerate ground states  $\ket{g_m}$ and excited states $\ket{e_m}$, where $m=\pm 1/2$ denotes the azimuthal quantum number with respect to the quantization axis $z$. The atom  interacts through its magnetic moment $\hat{\vec{\mu}}$ with a classical magnetic field $\vec{B}=(B_x,B_y,B_z)$, which drives a Larmor precession of the ground state atomic spin. The atom is coupled to an off-resonant linearly polarized laser field. This field, which is linearly polarized along the $x$-axis, can be decomposed into two circular components with annihilation operators $\hat a_{\pm} = (\mp \hat a_x + i \hat a_y)/\sqrt{2}$. Due to the dipole selection rules, the two circularly polarized field components  couple individually to the two different ground state populations.

During off-resonant probing (i.e., $\Delta\gg g$), the atomic excited state can be adiabatically eliminated, and the quantized field-atom Hamiltonian is given by \cite{Nielsen2008}:
\begin{align}
\label{eq:HamPM}
\hat H = \frac{\hbar g^2}{\Delta}\sum_{\ell = \pm 1}
\hat a_{\ell}^{\dag} \hat a_{\ell}\ket{g_{\ell/2}}\bra{g_{\ell/2}} + \hat{\vec{\mu}}\cdot\vec{B}.
\end{align}
In Eq.~(\ref{eq:HamPM}), $\Delta=\omega_L-\omega_A$ is the laser atom detuning, $g = \vec{d}\cdot\vec{E}_0/\hbar$, where $\vec{d}$
is the atomic electric dipole moment and $|\vec{E}_0| = \sqrt{\hbar\omega_L/(V\epsilon_0)}$ is the electric
field per photon in the quantization volume $V$, and $\epsilon_0$ is the (electric) vacuum permeability.

As as result of Eq.~(\ref{eq:HamPM}), the two circularly polarized field components experience phase shifts that depend on the atomic occupation of the two ground states. The resulting phase difference between the two field components implies a (Faraday) rotation of the field polarization, which is proportional to the population difference between the ground states $\ket{g_{\pm 1/2}}$.

Because of the strong
linearly polarized probe field with photon number $N_{\mathrm{ph}}\gg 1$,
the Stokes operators of the field can be written as

\begin{align}
\label{Stokes}
\hat J_x &= \frac{\hat a_x^{\dag} \hat a_x - \hat a_y^{\dag} \hat a_y}{2} \approx \frac{N_{\mathrm{ph}}}{2},\nonumber\\
\hat J_y &= \frac{\hat a_x^{\dag} \hat a_y + \hat a_y^{\dag} \hat a_x}{2}
\simeq \frac{\sqrt{N_{\mathrm{ph}}}}{2} (\hat a_y + \hat a_y^{\dag}) = \sqrt{N_{\mathrm{ph}}}\,\hat y,\nonumber\\
\hat J_z &= \frac{\hat a_x^{\dag} \hat a_y - \hat a_y^{\dag} \hat a_x}{2i}
\simeq \frac{\sqrt{N_{\mathrm{ph}}}}{2i} (\hat a_y - \hat a_y^{\dag}) = \sqrt{N_{\mathrm{ph}}}\,\hat p_y,\end{align}
defining the canonical conjugate operators $\hat{y}$ and $\hat{p}_y$. The polarization rotation of the field is conveniently measured by subtracting the intensities of polarization components linearly polarized at $\pm 45$ degrees with respect to the incident field, and when this quantity is expressed in terms of the Stokes observables we recover an expression proportional to the operator $\hat{y}$.

By writing $\hat{\vec{\mu}}=\mu \hat{\vec{\sigma}}$, where $\hat{\vec{\sigma}}=(\hat\sigma_x,\hat\sigma_y,\hat\sigma_z)$ are the Pauli matrices, and using the definitions in Eq.~(\ref{Stokes}), the total Hamiltonian can be written

\begin{align}
\label{eq:HamXY}
\hat H = \frac{\hbar g^2}{\Delta}\sqrt{N_{\mathrm{ph}}}\hat p_y\hat{\sigma}_z
+ \mu \sum_{\alpha = x,y,z}\hat\sigma_{\alpha} B_{\alpha}.
\end{align}
A term $ \frac{\hbar g^2}{2\Delta}(\hat a_x^{\dag} \hat a_x + \hat a_y^{\dag} \hat a_y)$
has been omitted from Eq.~(\ref{eq:HamXY}) as it gives rise to a common phase shift, but no polarization rotation of the probe laser field.

To properly account for the entanglement created between the atom and the continuous probe field, we treat the laser beam as a sequence of segments of length $L = c \tau$, area $A=V/L$ and volume $V$, each initially prepared in a coherent state before the interaction. The continuous interaction between the atom and the light beam can then be represented as a sequence of interactions of the atom with one segment after the other. Each segment, in turn, is described as a single harmonic oscillator mode described by the operators in Eq.~(\ref{Stokes}). In the time interval $\tau$ the atom interacts with $N_{\mathrm{ph}}=\Phi \tau$ photons, where $\Phi$ is the photon flux, and we assume that the interaction is sufficiently weak, that
the dynamics can be well described by the coarse-grained propagator
\begin{align}
\label{eq:propag}
\hat U = e^{-\frac{i}{\hbar}\hat H_{\tau}} \simeq e^{-\frac{i}{\hbar}\kappa_{\tau}\hat{p}_y\,\hat{\sigma}_{\vec{m}}}
e^{-\frac{i}{\hbar}\mu_{\tau}\vert\vec{B}\vert\hat\sigma_{\vec{n}_B}},
\end{align}
where
\begin{align}
\kappa_{\tau} = \frac{\hbar g^2\tau}{\Delta}\sqrt{N_{\mathrm{ph}}}, \qquad
\mu_{\tau} = \mu \tau,
\end{align}
and $\vec{n}_B\equiv (n_B^x,n_B^y,n_B^z)$ is
the unit vector pointing in the magnetic field direction. We have introduced $\hat{\sigma}_{\vec{n}_B} = \hat{\vec{\sigma}}\cdot\vec{n}_B$ and, in the part describing the atom-light interaction, we have introduced $\hat{\sigma}_{\vec{m}} = \hat{\vec{\sigma}}\cdot\vec{m}$, for the Pauli operator along an arbitrary unit vector direction
$\vec{m}$. To this aim we assume that we can probe the atomic ground state spin along any such direction with probe beams of appropriate polarization and propagation direction, or by application of a unitary spin rotation prior to probing of the $z$-component with a fixed beam set-up.
Since $N_{\mathrm{ph}}\propto\tau$ and $g\propto\tau^{-1/2}$ (through the volume
$V = A \tau c$), the dimensionless coupling constant $\kappa_{\tau}$ is proportional to $\tau^{1/2}$,
while $\mu_{\tau}$ is linear in $\tau$, and in Eq.~(\ref{eq:propag}), we have thus neglected terms of order $\tau^{3/2}$ and higher.

The incident laser beam is in a coherent state of linearly polarized light described by a Gaussian wave function $\pi^{-1/4}e^{-p_y^2/2}$ in the argument $p_y$, associated with the observable $\hat{p}_y$ introduced in Eq.~(\ref{Stokes}). The joint state of an atomic ground superposition state $\ket{\psi_{\mathrm{s}}(t)}=\sum_{\ell = \pm 1}c_{\ell/2}\ket{g_{\ell/2}}$ and the incident  $y$-polarization component of the quantized probe field can thus be written in the product basis $|p_y,g_{\ell/2}\rangle \equiv |p_y\rangle \otimes |g_{\ell/2}\rangle$,

\begin{align}
\ket{\Psi_{\mathrm{ps}}(t)} = \frac{1}{\pi^{1/4}}\sum_{\ell = \pm 1}
c_{\ell/2} \int\mathrm{d}p_y e^{-p^2_y/2} \ket{p_y,g_{\ell/2}}.
\end{align}
Under the action of (\ref{eq:propag}), this state evolves into the entangled state

\begin{align}
\label{eq:Evolstate}
\ket{\Psi_{\mathrm{ps}}(t+\tau)} &=  e^{-\frac{i}{\hbar}\mu_{\tau}\vert\vec{B}\vert\hat\sigma_{\vec{n}_B}}
e^{-\frac{i}{\hbar}\kappa_{\tau}\hat{p}_y\,\hat{\sigma}_{\vec{m}}}\ket{\Psi_{\mathrm{ps}}(t)}\nonumber\\
\phantom{=}&=\frac{1}{\pi^{1/4}} \sum_{\ell = \pm 1}
\int\mathrm{d}p_y \,c^{\prime}_{\ell/2}(p_y) e^{-p^2_y/2 } \ket{p_y,g_{\ell/2}}.
\end{align}
where, to first order in $\tau$ (second order in $\sqrt{\tau}$), the new expansion coefficients $c^{\prime}_{\ell/2}(p_y)$ are: 
\begin{align}
c^{\prime}_{\ell/2} &= c_{\ell/2}\!\!\left[
1 -  \frac{1}{2}\left(\frac{\kappa_{\tau}}{\hbar}\right)^2 \!\!p_y^2 - i\ell\left(
m_z\frac{\kappa_{\tau}}{\hbar} p_y +
n_B^z \frac{\mu_{\tau}\vert\vec B\vert}{\hbar}
\right)
\right]\nonumber\\
\phantom{=}&
-ic_{-\ell/2}\left[
\frac{\kappa_{\tau}}{\hbar} p_y
(m_x-i\ell m_y)
+\frac{\mu_{\tau}\vert\vec B\vert}{\hbar}
(n_B^x-i\ell n_B^y)
\right].
\end{align}

\subsection{A quantum filtering equation for the two-state atom}

To describe the back-action due to the field measurement, it is convenient to transform the entangled state (\ref{eq:Evolstate}) to the $y$ rather than $p_y$ representation of the field, using the relation

\begin{align}
\ket{p_y}=\frac{1}{\sqrt{2\pi}}\int\mathrm{d}y \,e^{-iyp_y}\ket{y}
\end{align}
and the state (\ref{eq:Evolstate}) can be rewritten as:

\begin{align}
\label{eq:EvolstateMes}
\ket{\Psi_{\mathrm{ps}}(t+\tau)} = \frac{1}{\pi^{1/4}}\sum_{\ell = \pm 1}
\int\mathrm{d}y \,\tilde{c}_{\ell/2}(y) e^{-y^2/2 } \ket{y,g_{\ell/2}}
\end{align}
with the new coefficient $\tilde{c}_{\ell/2}(y)$ given by

\begin{align}
\tilde{c}_{\ell/2} &=
ic_{-\ell/2}\left[
\frac{\mu_{\tau}\vert\vec B\vert}{\hbar}
(i\ell n_B^y-n_B^x)
+\frac{\kappa_{\tau}}{\hbar} y
(\ell m_y+im_x)
\right]
\nonumber\\
\phantom{=}&
\!\!\!\!\!\!\!\!\!\!+c_{\ell/2}\!\!\left[
1 \!- \! \frac{1}{2}\left(\frac{\kappa_{\tau}}{\hbar}\right)^2 \!\!(1 - y^2) - i\ell\!\left(
\!n_B^z \frac{\mu_{\tau}\vert\vec B\vert}{\hbar}
-im_z\frac{\kappa_{\tau}}{\hbar} y
\right)
\right].
\end{align}

A measurement of the light probe observable $\hat y$ with outcome $y^D$ projects the state of the
system (\ref{eq:EvolstateMes}) onto the state component with that definite value, \emph{i.e.}, the atomic part of the system becomes
\begin{align}
\label{eq:condatomstate}
\ket{\psi_{\mathrm s}(t+\tau)} =
\frac{1}{\pi^{1/4}}\sum_{\ell = \pm 1}\,\tilde{c}_{\ell/2}(y^D) e^{-\frac{(y^D)^2}{2}} \ket{g_{\ell/2}},
\end{align}
where the explicit dependence of $\tilde{c}_{\ell/2}$ on the measurement outcome $y^D$ causes an (infinitesimal) transfer of amplitude among the two atomic states.

Given the quantum state (\ref{eq:EvolstateMes}), the probability to measure a given value $y^D$ is
\begin{align}
\mathcal{P}(y^D) = \frac{e^{-(y^D)^2}}{\sqrt{\pi}}\sum_{\ell = \pm 1} \vert\tilde{c}_{\ell/2}(y^D)\vert^2
\simeq \pi^{-1/2} e^{-(y^D-y_0)^2},
\end{align}
where $y_0 = \frac{\kappa_{\tau}}{\hbar}\ave{\hat\sigma_{\vec{m}}}$. This explicitly shows how the optical probing yields a signal proportional to the desired mean value $\langle \hat{\sigma}_{\vec{m}}\rangle$, and it allows us to to
model the measurement outcome $y^D$ as a stochastic variable:

\begin{align}
\label{eq:yD}
y^D =  \frac{\kappa_{\tau}}{\hbar}\ave{\hat\sigma_{\vec{m}}} + \frac{\Delta W}{\sqrt{2\tau}},
\end{align}
where $\Delta W$ is a (finite) Gaussian Wiener increment with mean zero and variance $\tau$.

By replacing $y$ with the expression~(\ref{eq:yD}) for $y^D$ in
Eq.~(\ref{eq:condatomstate}), and by expanding the expressions for the state amplitudes to lowest order in $\tau$, we obtain, in the continuous limit, the quantum filtering equation for the state of the atomic system

\begin{align}
\mathrm d \hat\varrho_{\mathrm s} &= -i\frac{\mu_B}{\hbar}
[\hat{\sigma}_{\vec B},\hat\varrho_{\mathrm s}]
\mathrm d t
+ M \mathcal{D}[\hat\sigma_{\vec{m}}]\hat\varrho_{\mathrm s}\mathrm d t\nonumber\\
\phantom{=}&
+ \sqrt{M}\mathcal{H}[\hat\sigma_{\vec{m}}]\hat\varrho_{\mathrm s}\mathrm d W(t).
\label{eq:smeB}
\end{align}
Here the interaction parameter $M$ is given explicitly by

\begin{align}
M = \frac{g^4\tau^2}{4(\omega_L - \omega_A)^2}\Phi,
\end{align}
and an infinitesimal Wiener process models the noise in the detected signal, d$Y^D(t)=2\sqrt{M}\langle \hat{\sigma}_m\rangle \mathrm{d}t + \mathrm{d}W$. Thus, the system evolves according to a stochastic equation of the same form as the standard quantum filter equation (\ref{eq:sme1}).

We note that several probe fields may be  used to probe different spin components, and, to lowest order in $\tau$,  their effects on the quantum state commute. They may hence be included as separate terms with independent Wiener processes d$W_n$.

The modifications in Eq.~(\ref{eq:sme1}) due to finite detector efficiency can also be understood from first principles in the model system, and lead to similar correction factors in Eq.~(\ref{eq:smeB}).

\section{Conditional dynamics and estimation of a classical parameter}
\label{sec:two}

We are interested in the use of quantum systems to estimate a classical physical parameter or a set of parameters $\vec \gamma$. Here, in order to keep the discussion as general as possible, we treat the unknown
parameter $\vec\gamma$ as a vector quantity to indicate that it may be a set of parameters such as a damping rates, energy shifts, and coupling strengths, or, as in our example below, the directional components of a vector magnetic field. The experiment is sensitive to the value of these parameters, e.g., if they are coefficients in the Hamiltonian $\hat H = \hat H(\vec\gamma)$ acting on the system, and if this dependence results in a change of the observables probed in the experiment.

We describe the quantum dynamics of the combined system with $\vec\gamma$ belonging to a finite set of values
$\cal{V_{\gamma}}$$=\{\vec\gamma_k: k=1,\dots,N\}$.

For an observer who knows the true value of $\vec{\gamma}=\vec{\gamma}_{k_0}$, the system is described by our original reduced system stochastic master equation (\ref{eq:sme1}) with $H=H(\vec{\gamma}_{k_0})$,

\begin{align}
\label{eq:smek0}
\mathrm d \hat\varrho_{0}^{\mathrm{s}} &= \frac{i}{\hbar}
[\hat\varrho_{0}^{\mathrm{s}},\hat H(\vec{\gamma}_{k_0})]\mathrm d t
+ \sum_j\Gamma_j \mathcal{D}[\hat{\cal O}_j]\hat\varrho_{0}^{\mathrm{s}}\mathrm d t
\nonumber\\
\phantom{=}&
+ \sum_{n=1}^{M}M_n \mathcal{D}[\hat{\cal M}_n]\hat\varrho_{0}^{\mathrm{s}}\mathrm d t
+\sqrt{\eta_nM_n}
\mathcal{H}[\hat{\mathcal{M}}_n]\hat\varrho_{0}^{\mathrm{s}}
\mathrm{d}W_n(t),
\end{align}
conditioned by the measurement signals

\begin{align}
\label{eq:dY0}
\mathrm{d}Y_n^D(t)=\sqrt{\eta_n}\mathrm{d}W_n(t)
+\eta_n\sqrt{M_n}\ave{\hat{\cal{M}}_n+\hat{\cal{M}}_n^{\dagger}}_{0}\mathrm{d}t.
\end{align}
where $\ave{\hat{\cal{M}}_n+\hat{\cal{M}}_n^{\dagger}}_{0}=\mathrm{Tr}_S\{(\hat{\cal{M}}_n+
\hat{\cal{M}}_n^{\dagger})\hat\varrho_{0}^{\mathrm{s}}\}$,
and where d$W_n(t)$ are standard Wiener processes.
Here $\mathrm{Tr}_S(\cdot)$ is the trace on the system Hilbert space.

In the following we will assume that we probe all three spin components $(\hat{\sigma}_x,\hat{\sigma}_y,\hat{\sigma}_z)$ with measurement strengths $(M_1,M_2,M_3)$ and efficiencies $(\eta_1,\eta_2,\eta_3)$. By introducing the three-dimensional real Bloch vector, $\vec{r}= \mathrm{Tr}(\varrho_{0}^{\mathrm{s}}\hat{\vec{\sigma}})$, the stochastic master equation (\ref{eq:smek0}) can be rewritten in the Bloch vector representation,
\begin{align}
\label{eq:bloch1}
\mathrm{d}\vec{r} &= 2\left\{
\left[(\vec{b}_{k_0}\times\vec{r})-(\alpha_1+\alpha_2+\alpha_3)\vec{r}+\vec{\alpha}*\vec{r}
\right]\right.
\nonumber\\
\phantom{=}&+\left.
(1-r^2\mathrm{d}\vec{\Omega})-\vec{r}\times(\vec{r}\times\mathrm{d}\vec{\Omega})
\right\},
\end{align}
where we have passed to dimensionless units by performing the replacement $t\rightarrow Mt$ with $M=\max\{M_1,M_2,M_3\}$. Besides this, we have defined the vectors $\vec\alpha$, with components $\alpha_n=M_n/M$, $\vec{\eta}=(\eta_1,\eta_2,\eta_3)$, $\vec{b}_k=\mu_B\vec{B}_k/(\hbar M)$, and $\mathrm{d}\vec{\Omega}$, with components d$\Omega_n =  \sqrt{\eta_n\alpha_n}\mathrm{d}W_n$. The symbol $*$
in Eq.~(\ref{eq:bloch1}) indicates the pointwise product, $(\vec{p}*\vec{q})_n \equiv p_nq_n$.

Equation~(\ref{eq:bloch1}) was derived and analyzed in detail in Refs.~\cite{Ruskov2010,Ruskov2012}, for the special case of identical probing strengths and efficiencies for all three orthogonal spin directions. As shown in that work, in the absence of a magnetic field, the Bloch vector is driven towards a steady state radial (purity) distribution and an isotropic angular distribution within the Bloch sphere. The magnetic field, in turn, provides a torque for the atomic spin vector, and the resulting spin precession, in the plane perpendicular to the field, reveals itself in a modulation of the polarization rotation measurements.

\subsection{Augmented quantum filter equation}

In our formulation of the estimation problem we will treat the classical parameter $\vec\gamma$ as unknown with an assigned probability distribution $P(\vec{\gamma})$. The measurements then cause an update of the probability distribution, which is governed by Bayes rule for conditional probabilities. Until the actual value of $\vec{\gamma}$ is known, we thus have to treat each candidate value with a probability factor, and for each possible value of $\vec{\gamma}$, the corresponding state of the quantum system evolves under the quantum filtering equation with the corresponding dependence of $\vec{\gamma}$.

To this end it is convenient~\cite{Molmer2004,Gambetta2005,Chase2009,Ralph2011,Tsang2012} to consider the augmented Hilbert space
$\mathfrak{H}=\mathfrak{H}_{\mathrm{s}}\otimes\mathfrak{H}_{\gamma}$, where $\mathfrak{H}_{\mathrm{s}}$ and $\mathfrak{H}_{\gamma}$ refer to
the quantum system Hilbert space and the space of classical states for the variable $\vec\gamma$, respectively. The latter space describes states with definite values of the parameters, and superposition states are not populated. The quantum mechanical notation, however, still applies and, e.g., describes a probability distribution for a set of values $\vec{\gamma}_k$ as a diagonal density matrix $\hat\varrho_{\gamma}=\sum_k P_k \ket{\vec\gamma_k}\bra{\vec\gamma_k}$. The classical variables $\vec{\gamma}$ are equivalent to quantum non-demolition (QND) variables of an ancillary quantum system that interacts with our probe system, and for which the Bayesian probability update is fully equivalent to the quantum measurement back-action. When we incorporate the parameters $\vec{\gamma}$ in this way we can directly apply the filtering equation on the augmented space.

The observer who does not know the value of $\vec{\gamma}$ describes
the combined quantum and classical system by the augmented density matrix
\begin{align}
\label{eq:statecom}
\hat\varrho = \sum_{k=1}^N P_k \ket{\vec{\gamma}_k}\bra{\vec{\gamma}_k}\otimes\hat\varrho_k^{\mathrm{s}},
\end{align}
where $\hat\varrho^{\mathrm{s}}_k$ is the normalized system state density matrix associated with the specific value $\vec{\gamma}=\vec{\gamma}_k$.

The combined system evolves according to the quantum filter equation
\begin{align}
\label{eq:sme}
\mathrm d \hat\varrho &= \left(\frac{i}{\hbar}
[\hat\varrho,\hat H]
+ \sum_j\Gamma_j \mathcal{D}[\hat{\cal O}_j]\hat\varrho
+ \sum_{n=1}^{M}M_n \mathcal{D}[\hat{\cal M}_n]\hat\varrho\right)\mathrm d t
\nonumber\\
\phantom{=}&\!\!\!\!\!\!\!\!
+ \!\!\sum_{n=1}^{M}\!\sqrt{M_n}\mathcal{H}[\hat{\cal M}_n]\hat\varrho
\left(\!
\mathrm{d}Y_n^D(t)-\eta_n\sqrt{M_n}\ave{\hat{\cal{M}}_n+\hat{\cal{M}}_n^{\dagger}}_E\mathrm{d}t\!
\right)\!,
\end{align}
where the operator $\mathcal{H}$ is defined as:

\begin{align}
\mathcal{H}[\hat f]\hat\varrho &= \hat f \hat\varrho + \hat\varrho \hat f^{\dag}
- \ave{\hat f + \hat f^{\dag}}_E \hat\varrho.
\label{eq:superop}
\end{align}
Here we use the notation $\ave{\hat{f}}_E=\mathrm{Tr}(\hat{f}\hat{\varrho})=\sum_{k=1}^NP_k\mathrm{Tr}_S(\hat f\hat\varrho_k^{\mathrm{s}})$ to explicitly recall that the expectation value of the signal should be determined by the full augmented quantum state, equivalent to a weighted average over the ensemble of states of the quantum system governed by the different values of $\vec{\gamma}$.

If the dependence on $\vec{\gamma}$ only enters via the Hamiltonian $\hat{H}(\vec{\gamma})$, the different terms in (\ref{eq:sme}) are implemented as the following product operators on  $\mathfrak{H}=\mathfrak{H}_{\mathrm{s}} \otimes\mathfrak{H}_{\gamma}$

\begin{align}
\hat H &= \sum_{k=1}^N\ket{\vec{\gamma}_k}\bra{\vec{\gamma}_k}\otimes\hat H(\vec\gamma_k),
\end{align}
\begin{align}
\hat{\mathcal{O}}_j &\equiv \sum_{k=1}^N\ket{\vec{\gamma}_k}\bra{\vec{\gamma}_k}\otimes\hat{\mathcal{O}}_j,\\ \nonumber
\hat{\mathcal{M}}_n &\equiv \sum_{k=1}^N\ket{\vec{\gamma}_k}\bra{\vec{\gamma}_k}\otimes\hat{\mathcal{M}}_n.
\end{align}

\subsection{Detection signal properties}

The stochastic process appearing in Eq.(\ref{eq:sme}),

\begin{align}
\label{eq:dvn}
\sqrt{\eta_n}\mathrm{d}V_n(t):= \mathrm{d}Y_n^D(t)-\eta_n\sqrt{M_n}\ave{\hat{\cal{M}}_n+\hat{\cal{M}}_n^{\dagger}}_E\mathrm{d}t.
\end{align}
is not a standard Wiener process. This is because we subtract from the measured signal a weighted average based on our probabilistic description of $\vec{\gamma}$, while the measured photocurrent d$Y^D_n$ in the experiment is governed by the actual value of the unknown parameter $\vec{\gamma}=\vec{\gamma}_{k_0}$.

Equation~(\ref{eq:dY0}) characterizes the properties of such a realistic detection record, and, when inserted in Eq.~(\ref{eq:sme}), we find that the stochastic process in Eq.~(\ref{eq:dvn}) can be rewritten as

\begin{align}
\label{eq:dvn-bis}
\mathrm{d}V_n \!= \!\mathrm{d}W_n\! - \!\sqrt{\eta_n M_n}
\!\left[\!
\ave{\hat{\cal{M}}_n+\hat{\cal{M}}_n^{\dagger}}_E-\ave{\hat{\cal{M}}_n+\hat{\cal{M}}_n^{\dagger}}_{0}\!
\right]\!
\mathrm{d}t.
\end{align}
Hence, d$V_n(t)$ is a stochastic Gaussian process with variance $\mathrm{d}t$,
and with (statistical) mean value $\langle\langle \mathrm{d}V_n(t)\rangle\rangle =\sqrt{\eta_n M_n}(\ave{\hat{\cal{M}}_n+\hat{\cal{M}}_n^{\dagger}}_{0}-\ave{\hat{\cal{M}}_n+\hat{\cal{M}}_n^{\dagger}}_E)\mathrm{d}t$, reflecting precisely the difference in the expectation value assumed by the weighted average over different $\vec{\gamma}_k$ and by the correct value $\vec{\gamma}_{k_0}$.

Equation~(\ref{eq:sme}) permits a full simulation of the detection process and provides a time dependent solution of the form
\begin{align}
\label{eq:statecom2}
\hat\varrho = \sum_{k=1}^N \ket{\vec{\gamma}_k}\bra{\vec{\gamma}_k}\otimes\hat\rho_k^{\mathrm{s}},
\end{align}
where each value of $\vec{\gamma}_k$ is associated with an unnormalized $\hat{\rho}_k^s$. Equation~ (\ref{eq:statecom2}) is of course equivalent to Eq.~(\ref{eq:statecom}) with the normalized system density matrix $\hat{\varrho}_k^s$ and the probability distribution $P_k$.

As shown in detail in the appendix~\ref{sec:apA}, Eq.~(\ref{eq:sme}) leads to two separate equations for $\hat{\varrho}_k^s$ and $P_k$:

\begin{align}
\label{eq:smekntris}
\mathrm d \hat\varrho_{k}^{\mathrm{s}} &= \frac{i}{\hbar}
[\hat\varrho_{k}^{\mathrm{s}},\hat H(\vec{\gamma}_{k})]\mathrm d t
+ \sum_j\Gamma_j \mathcal{D}[\hat{\cal O}_j]\hat\varrho_{k}^{\mathrm{s}}\mathrm d t
\nonumber\\
\phantom{=}&
+ \sum_{n=1}^{M}M_n \mathcal{D}[\hat{\cal M}_n]\hat\varrho_{k}^{\mathrm{s}}\mathrm d t
+\sqrt{M_n}
\mathcal{H}[\hat{\mathcal{M}}_n]\hat\varrho_{k}^{\mathrm{s}}
\mathrm{d}\tilde{V}^{(k)}_n(t),
\end{align}
with d$\tilde{V}^{(k)}_n(t)=\mathrm{d}Y_n^D(t)-\eta_n\sqrt{M_n}\ave{\hat{\cal{M}}_n+\hat{\cal{M}}_n^{\dagger}}_k\mathrm{d}t$, and

\begin{align}
\label{eq:newdPk}
\mathrm{d}P_k&=\mathrm{Tr}_S(\mathrm d \hat\rho_k^{\mathrm{s}})\nonumber\\
\phantom{=}&\!\!\!\!\!\!\!\!\!\!\!\!
=P_k\!\sum_{n=1}^{M}\!\sqrt{\eta_nM_n}\left[
\ave{\hat{\cal{M}}_n+\hat{\cal{M}}_n^{\dagger}}_k-\ave{\hat{\cal{M}}_n+\hat{\cal{M}}_n^{\dagger}}_E
\right]\!\mathrm{d}V_n(t).
\end{align}
In both equations the  photocurrent d$Y_n^D(t)$, observed or simulated according to Eq.~(\ref{eq:smek0}), appears, and
Eq.~(\ref{eq:newdPk}) agrees with the expression given in Ref.~\cite{Chase2009}. We note, however, that the stochastic 
term in our Eq.~(\ref{eq:smekntris}) contains the expectation value $\ave{\hat{\cal{M}}_n+\hat{\cal{M}}_n^{\dagger}}_k$ 
corresponding to the parameter value $\vec{\gamma}_k$, while in Ref.~\cite{Chase2009} the ensemble average 
$\ave{\hat{\cal{M}}_n+\hat{\cal{M}}_n^{\dagger}}_E$ has been used.

By inserting the expression (\ref{eq:dvn-bis}) for  d$V_n$ in Eq.~(\ref{eq:newdPk}), we see that the
change of d$P_k$ due to the measurements is given by

\begin{widetext}
\begin{align}
\left(
\ave{\hat{\cal{M}}_n+\hat{\cal{M}}_n^{\dagger}}_k-\ave{\hat{\cal{M}}_n+\hat{\cal{M}}_n^{\dagger}}_E
\right)
\left[
\mathrm{d}W_n(t)+\sqrt{\eta_nM_n}\left(
\ave{\hat{\cal{M}}_n+\hat{\cal{M}}_n^{\dagger}}_{0}-\ave{\hat{\cal{M}}_n+\hat{\cal{M}}_n^{\dagger}}_E
\right)\mathrm{d}t
\right].
\end{align}
\end{widetext}

This equation has a natural interpretation: For parameter values $\vec\gamma_k$ where the expected mean current $\propto \ave{\hat{\cal{M}}_n+\hat{\cal{M}}_n^{\dagger}}_k$ differs in the same (opposite) direction from the ensemble mean as the one expected for the actual value $\ave{\hat{\cal{M}}_n+\hat{\cal{M}}_n^{\dagger}}_{0}$, the two parentheses will typically have the same sign, and $P_k$ will increase (decrease). Due to the random contribution $\mathrm{d}W_n(t)$, however, the probabilities will show fluctuations, and their increase (decrease) with time will appear only as an average trend, leading, in particular, to a typically increasing value for the probability of the correct value $P_{k_0}$.

\section{Vector magnetometry with a two-level atom}

We now apply the formalism to the two-level atom coupled to a magnetic field $\vec{B}$ with known magnitude $|\vec{B}|$, but unknown direction in space. Since such a field will cause a spin precession around the magnetic field axis, we expect that optical probing of a single spin component will not be sensitive to the magnetic field projection along the spin direction probed, while the simultaneous probing of all three spin components is bound to reveal any motion of the mean spin vector due to the  magnetic precession.


The acquisition of data from a real or a simulated experiment cause a continuous update of the probability distribution $P_k$ for the magnetic field, represented in the following by the dimensionless vector, $\vec{b}_k=\mu_B\vec{B}_k/(\hbar M)$. The angular measure, $\cos\theta(t) = \vert\vec{b}_u\vert^{-2}\sum_k P_k(t)\,(\vec{b}_k\cdot\vec{b}_u)$, quantifies the scattering of the magnetic field directions $\vec{b}_k$ inferred from a single experimental run around the actual value $\vec{b}_u$. By carrying out a large number of simulations, we thus quantify the average performance of the method by the (average) scalar product
\begin{align}
\label{eq:cosE}
\langle\langle\cos\theta\rangle\rangle(t) = \vert\vec{b}_u\vert^{-2}
 \sum_k \left\langle\left\langle P_k(t)\,(\vec{b}_k\cdot\vec{b}_u)
 \right\rangle\right\rangle
\end{align}
as a function of the measurement time $t$.

\subsection{Direction of a magnetic field along a given axis}

\begin{figure}[t]
\begin{center}
\includegraphics{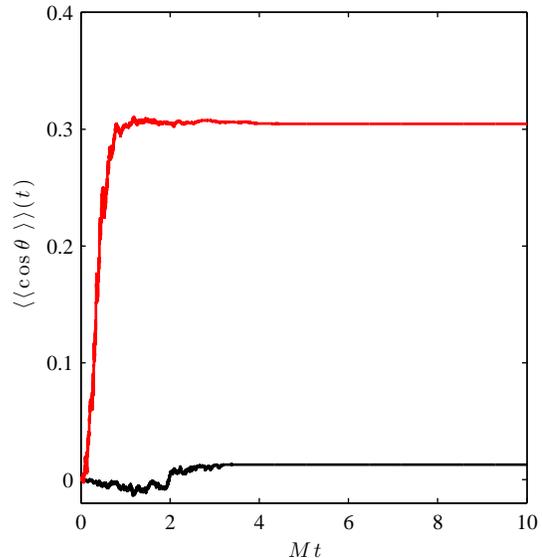}
\end{center}
\caption{(Color online). Time evolution of $\langle\langle\cos\theta\rangle\rangle(t)$ for $\vec{b}_u=(0.1,0,0)$ (lower, black line)
and $\vec{b}_u=(1.5,0,0)$ (upper, red line) for the qubit example discussed in Ref.~\cite{Chase2009}.}
\label{fig:one}
\end{figure}

Following Ref.~\cite{Chase2009}, we have first investigated the case in which the initial state of the atom is represented by the Bloch vector $\vec{r}_0=(0,1,0)$ with $M_{1,2}=0$ but $M_3\ne 0$ (i.e., we probe only the component of the spin along $z$ and $M\equiv M_3$), and where the magnetic field has a known strength while it has equal prior probabilities to point in the positive and in the negative $x$-axis directions. In Fig.~\ref{fig:one} we show the behavior of $\langle\langle\cos\theta\rangle\rangle(t)$ as function of time in the two cases of a weak $\vert\vec{b}_u\vert = 0.1$ (lower, black curve) and a strong $\vert\vec{b}_u\vert = 1.5$ field (upper, red curve).
The curves are obtained by averaging over 104 simulated detection records. The stronger the field amplitude, the better is the directional estimate, but, as also observed in Ref.~\cite{Chase2009}, the quantum filter does not unambiguously identify the direction $\vec{b}_u$ of the unknown field.

\begin{figure}[t]
\begin{center}
\includegraphics{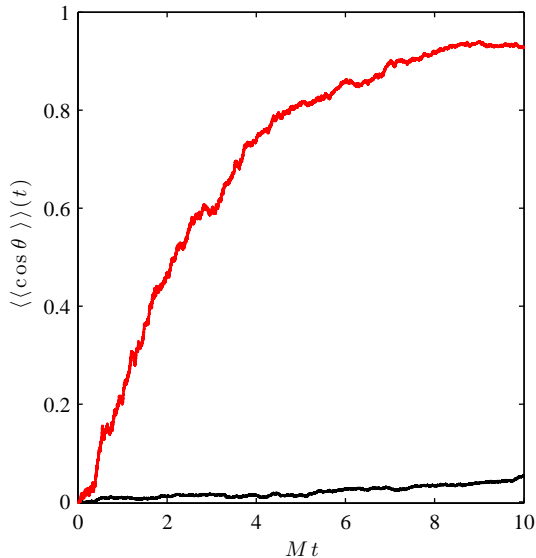}
\end{center}
\caption{(Color online). Same setup as in Fig.~\ref{fig:one}, but with three detectors ($M\equiv M_1=M_2=M_3$). }
\label{fig:two}
\end{figure}

This situation changes when all the three spin components are detected. As illustrated in Fig.~\ref{fig:two}, the accumulation of results from all three detectors lead, especially for high strength of the field, to a final probability distribution which is well converged to the correct $\vec{b}_u$. In these numerical experiments 104 trials have been performed in order to accumulate sufficient data for the statistical
averages within a reasonable computational time. A small fraction ($\sim\,1$\%) of the simulated trajectories are unstable in the case of the three detectors and they have been rejected from the statistical averages.

There is a number of competing effects that may explain the dependence on the quality of our estimate on the field strength and the number of spin components probed: Measuring the $z$-component of a spin may lead to a (Zeno-effect) suppression of, and, hence, insensitivity to, slow precession of the spin around the $x$-axis, while even for strong fields, the measurement of a single spin component does not allow discrimination between left and right circular precession around the $x$-axis. The probing of several components on the one hand allow discrimination between left and right circular precession, and, on the other hand prevents (Zeno-) locking of the system to eigenstates of the probed quantities as they do not commute and do not have common eigenstates.


\subsection{Estimation of a spherically random direction of a magnetic field}

\begin{figure*}[htb!]
\begin{center}
\includegraphics{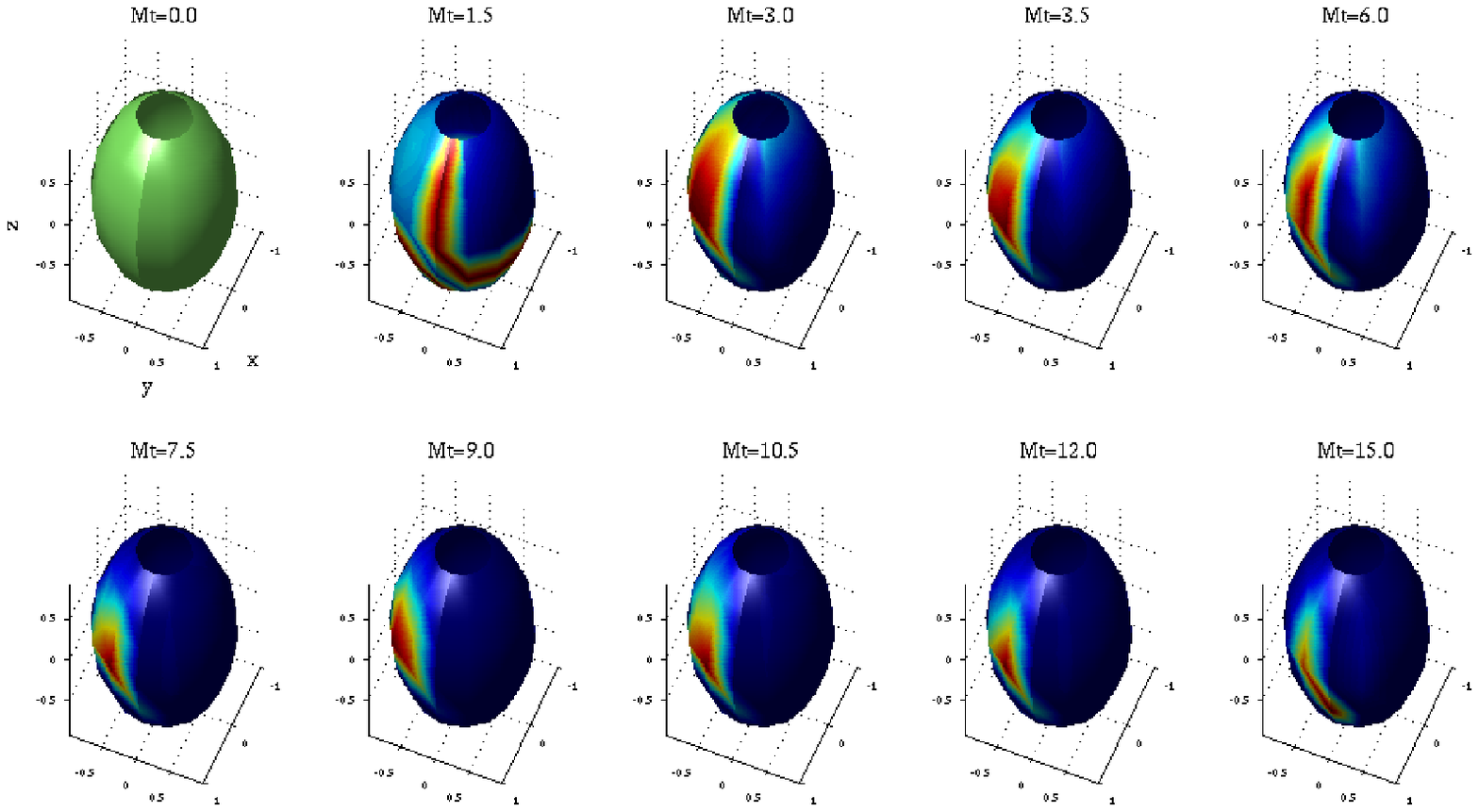}
\includegraphics{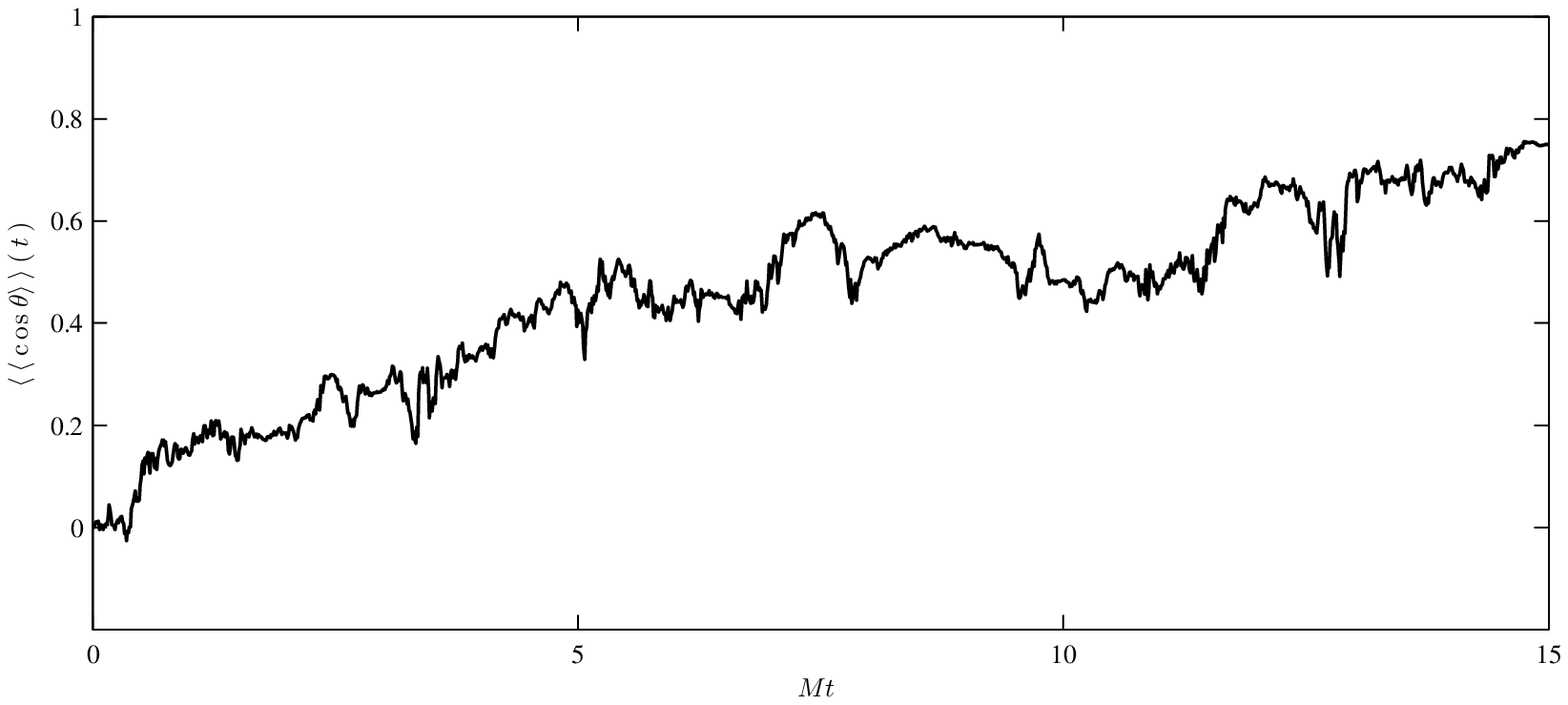}
\end{center}
\caption{(Color online). Upper panels: Time evolution of the probability distribution $P_k$ for an ensemble $\mathcal{V}_B$ of 98 elements initially
equally distributed over the unit sphere. The unknown magnetic field that has to be determied is the strong field $\vec{b}_u = (1.5,0,0)$, and 
its direction is displayed by the pink dot on the sphere at time $Mt= 15$. 
Lower panel: Evolution of $\langle\langle\cos\theta\rangle\rangle(t)$ for the single shot experiment simulated in the upper panels.
 }
\label{fig:four}
\end{figure*}

Finally, we have analyzed the scenario in which the magnetic field has an isotropic prior probability distribution, represented by an ensemble $\mathcal{V}_B$ with $N=98$ directions on the unit sphere, as
illustrated in Fig.~\ref{fig:four}. Here, we assume the stronger field $\vert\vec{b}_u\vert = 1.5$, and we assume equal strength probing of all three spin components.

The initial condition for the
probabilities with all $P_k=1/N$ is illustrated by the (green) sphere displayed in Fig.~\ref{fig:four} at time $Mt=0$. We study the convergence of the probability distribution as a function of the probing time, and for long times ($MT= 15$), the filter converges well to a single direction. We note that the spheres are cut on the top because in the simulation
we have considered an interval $\theta = (0,\pi)$ equally spaced with $N_{\theta}=7$ grid points and an interval $\phi=[0,2\pi)$ with $N_{\phi}=14$ grid points
for the azimuthal angle.

The equal strength probing of the three cartesian spin components $\sigma_x,\ \sigma_y,\ \sigma_z$ is equivalent ~\cite{Chase2009,Ruskov2010} to probing of any other cartesian set, including for example one parallel component and two perpendicular components to the applied magnetic field, and we find that the monitoring of all three spin components lead to unambiguous identification of the direction of the applied field. 

The spin vector may, inadvertently, align, parallel or anti-parallel with the applied magnetic field axis, but the random back-action of the probing of the non-commuting orthogonal spin observables will kick the system away from these directions and give rise to renewed observable precession. 

\section{Conclusion}

In this work, we have demonstrated a Bayesian filter for classical parameters, which affect the dynamics of a quantum system. Previous studies along the same lines have focused on quantum non-demolition measurements of typically a single variable, but as shown by our analysis, a non-QND setting may be analyzed by the very same assumptions and methods. Non-QND probing may have specific advantages and provide more decisive results, when the parameters affect different observables of the quantum probe, as illustrated explicitly by our numerical simulations.

The Bayesian filter is derived from a standard quantum filter formulation of conditioned quantum dynamics. In this mapping, we model the classical parameters as QND observables of auxiliary quantum systems, and their classical probability distribution thus coincides with the conventional reduced density matrix elements for a quantum system.
Since the quantum state description cannot be completed by further knowledge in the form of hidden variables,
our formulation of the parameter estimation problem, indeed, provides the tightest and most precise probability distribution for the variable of interest conditioned on the measurement outcome and on the prior probability distribution.

 The discretisation of the parameter space and solution of a quantum system master equation associated with each potential parameter value naturally puts limit on the precision of the method and the number of variables that can be realistically determined. A natural next step would be to apply methods that gradually refine the parameter space around the most likely values and suppress the most unlikely ones from the calculation. Such weighted stochastic differential equations are known in statistics~\cite{James2006}, and we imagine that they may be used to decide objective means to suppress and to breed new parameter values without enlarging the memory and computational demands of the method. Another interesting approach, put forward in Ref.~\cite{Nielsen2009}, involves projection of the complete system on a non-linear lower-dimensional manifold on which the integration of the stochastic differential equations of motion is faster. Alternatively, maximum likelihood methods and random searches through the parameter space, e.g., by Markov Chain Monte Carlo methods~\cite{Press2007}, may be effectively applied to even very large search spaces. We imagine that our simulations may serve as useful reference data for testing such alternative estimation techniques.

\section*{Acknowledgements}

This work was supported by the EU integrated project AQUTE (K.M.), the EU collaborative project QIBEC, the Marie-Curie Programme of the EU through
Proposal Nr. 236073 (OPTIQUOS) within the 7th European Community Framework Programme, the Deutsche Forschungsgemeinschaft within the Grant No. SFB/TRR21, 
and the excellence cluster 'The Hamburg Centre for Ultrafast Imaging - Structure, Dynamics and Control of Matter at the Atomic Scale' of 
the Deutsche Forschungsgemeinschaft (A.N.). A.N. acknowledges also the bwGrid for computational resources and Mr. J\"urgen Salk for his technical support. 

\section*{Appendix}

\subsection{Quantum filtering equations for the parameter estimation problem}
\label{sec:apA}

In order to derive the equations~(\ref{eq:smekntris}) and~(\ref{eq:newdPk}) we first
need the SME for $\hat\rho_k^{\mathrm{s}}$, which is given by

\begin{widetext}
\begin{align}
\label{eq:smeknn}
\frac{\mathrm d \hat\rho_k^{\mathrm{s}}}{P_k} \!&= \!\left(\frac{i}{\hbar}
[\hat\varrho_k^{\mathrm{s}},\hat H(\vec{\gamma}_k)]
+ \sum_j\Gamma_j \mathcal{D}[\hat{\cal O}_j]\hat\varrho_k^{\mathrm{s}}
+ \sum_{n=1}^{M}M_n \mathcal{D}[\hat{\cal M}_n]\hat\varrho_k^{\mathrm{s}}\right)\!\mathrm d t
+ \sum_{n=1}^{M}\!\sqrt{\eta_nM_n}\!\left(\hat{\mathcal{M}}_n\hat\varrho_k^{\mathrm{s}}
+\hat\varrho_k^{\mathrm{s}}\hat{\mathcal{M}}_n^{\dagger}
-\ave{\hat{\mathcal{M}}_n+\hat{\mathcal{M}}_n^{\dagger}}_E\hat\varrho_k^{\mathrm{s}}
\right)\!\mathrm{d}V_n(t),
\end{align}
\end{widetext}
where we note that $\mathrm d \hat\rho_k^{\mathrm{s}}\equiv\mathrm d (\bra{\vec{\gamma}_k}\hat\varrho\ket{\vec{\gamma}_k})$.
Given this,  the equation of motion for the probability $P_k$ is easily obtained by computing the trace of Eq.~(\ref{eq:smeknn}), which
provides precisely Eq.~(\ref{eq:newdPk}). Now, since $\hat\varrho_k^{\mathrm{s}}=\hat\rho_k^{\mathrm{s}}/P_k$ we have:

\begin{align}
\label{eq:itoformuladrho}
\mathrm{d}\hat\varrho_k^{\mathrm{s}}=
\frac{1}{P_k}\cdot\mathrm{d}\hat\rho_k^{\mathrm{s}}+\hat\rho_k^{\mathrm{s}}\cdot\mathrm{d}\left(\frac{1}{P_k}\right)
+\mathrm{d}\hat\rho_k^{\mathrm{s}}\cdot\mathrm{d}\left(\frac{1}{P_k}\right),
\end{align}
where (classical It\^o formula~\cite{Gardiner2004})

\begin{align}
\mathrm{d}\left(\frac{1}{P_k}\right)=-\frac{1}{P_k^2}\mathrm{d}P_k
+\frac{1}{P_k^3}(\mathrm{d}P_k)^2.
\end{align}
Thus, we have

\begin{align}
\frac{(\mathrm{d}P_k)^2}{P_k^2} = \sum_{n=1}^{M}\eta_nM_n\left(
\ave{\hat{\cal{M}}_n+\hat{\cal{M}}_n^{\dagger}}_k-\ave{\hat{\cal{M}}_n+\hat{\cal{M}}_n^{\dagger}}_E
\right)^2\!\mathrm{d}t,
\end{align}
where we used the fact that $\mathrm{d}W_n(t)\mathrm{d}t=0$ and $\mathrm{d}W_n(t)\mathrm{d}W_{n^{\prime}}(t)=\delta_{n,n^{\prime}}\mathrm{d}t$.
Consequently, we obtain

\begin{align}
\mathrm{d}\!\!\left(\!\frac{1}{P_k}\!\right)\!&=\!\frac{1}{P_k}
\!\sum_{n=1}^{M}\left\{\eta_nM_n\left(
\ave{\hat{\cal{M}}_n+\hat{\cal{M}}_n^{\dagger}}_k-\ave{\hat{\cal{M}}_n+\hat{\cal{M}}_n^{\dagger}}_E
\right)\right.
\nonumber\\
\phantom{=}&
\times
\left(
\ave{\hat{\cal{M}}_n+\hat{\cal{M}}_n^{\dagger}}_k-\ave{\hat{\cal{M}}_n+\hat{\cal{M}}_n^{\dagger}}_{k_0}
\right)\mathrm{d}t-\sqrt{\eta_nM_n}
\nonumber\\
\phantom{=}&\left.
\times \left(
\ave{\hat{\cal{M}}_n+\hat{\cal{M}}_n^{\dagger}}_k-\ave{\hat{\cal{M}}_n+\hat{\cal{M}}_n^{\dagger}}_E
\right)\mathrm{d}W_n(t)\right\},
\end{align}
and therefore

\begin{align}
\mathrm{d}\hat\rho_k^{\mathrm{s}}\cdot\mathrm{d}\!\!\left(\!\frac{1}{P_k}\!\right)&\!=\!
\sum_{n=1}^{M}\eta_nM_n\!\left(
\ave{\hat{\cal{M}}_n+\hat{\cal{M}}_n^{\dagger}}_E - \ave{\hat{\cal{M}}_n+\hat{\cal{M}}_n^{\dagger}}_k
\right)\nonumber\\
\phantom{=}&\times
\left(
\hat{\mathcal{M}}_n\hat\varrho_k^{\mathrm{s}}
+\hat\varrho_k^{\mathrm{s}}\hat{\mathcal{M}}_n^{\dagger}
-\ave{\hat{\mathcal{M}}_n+\hat{\mathcal{M}}_n^{\dagger}}_E\hat\varrho_k^{\mathrm{s}}
\right)\mathrm{d}t.
\end{align}
Putting all together into Eq.~(\ref{eq:itoformuladrho}) and by using the definition~(\ref{eq:dY0}) we derive (\ref{eq:smekntris}).

\subsection{Numerical simulations}

In analogy to Eq.~(\ref{eq:bloch1}), we can represent the (normalized) density matrices associated with each value $\vec{\gamma}_k$ as a Bloch vector, and propagate the collection of Bloch vectors as subject to the noisy detection signals - governed by Eq.~(\ref{eq:dY0}). Using the same notation as in Eq.~(\ref{eq:bloch1}), these equations have the form

\begin{align}
\label{eq:sderk}
\mathrm{d}\vec{r}_{k} &= 2\left\{
\left[(\vec{b}_{k}\times\vec{r}_{k})-(\alpha_1+\alpha_2+\alpha_3)\vec{r}_{k}+\vec{\alpha}*\vec{r}_k
\right.\right.
\nonumber\\
\phantom{=}&+\left.
(1-\delta_{k_0,k})\left(
\vec{{I}}_k(1-r_{k}^2)-\vec{r}_{k}\times(\vec{r}_{k}\times\vec{{I}}_k)
\right)
\right]\mathrm{d}t
\nonumber\\
\phantom{=}&+\left.
\mathrm{d}\vec{\Omega}(1-r_{k}^2)-\vec{r}_{k}\times(\vec{r}_{k}\times\mathrm{d}\vec{\Omega})
\right\},
\end{align}
Instead, the probabilities represented as the column vector $\vec{P}=(P_1,P_2,\dots,P_N)^{\mathsf{T}}$ obey the following matrix equation

\begin{align}
\mathrm{d}\vec{P} &= 4\left\{
\vec{P}*(\vec{\upsilon}^{\mathsf{T}}\cdot \mathbf{C})^{\mathsf{T}}
- \vec{P}\left[\vec{P}^{\mathsf{T}}\cdot(\vec{\upsilon}^{\mathsf{T}}\cdot \mathbf{C})^{\mathsf{T}})\right]\right.\nonumber\\
\phantom{=}&-
\left.\vec{P}*\left[(\mathbf{C}\,\vec{P})^{\mathsf{T}}\tilde{\mathbf{C}}\right]^{\mathsf{T}}
+\vec{P}\left[(\mathbf{C}\,\vec{P})^{\mathsf{T}}\cdot(\tilde{\mathbf{C}}\,\vec{P})\right]
\right\}\mathrm{d}t
+G\,\mathrm{d}\vec{\Omega}.
\label{eq:sdeP}
\end{align}
Here, $C$ and $\tilde{C}$ are $3\times N$ matrices and $G$ is an $N\times 3$ matrix containing the Bloch vector solutions of Eq.~(\ref{eq:sderk}), $C_{n,j}=x^{(n)}_j$, $\tilde{C}_{n,j}=\eta_n\alpha_nC_{n,j}$, and $G_{j,n}=2P_j(x_j^{(n)}-\vec{P}^{\mathsf{T}}\cdot\vec{X}_n)$ where $\vec{X}_n=(x_1^{(n)},x_2^{(n)},\cdots,x_N^{(n)})^{\mathsf{T}}$,
and $\vec{\upsilon}=\vec{\eta}*\vec{\alpha}*\vec{r}_{k_0}$.

For the numerical simulation of both~(\ref{eq:sderk}) and~(\ref{eq:sdeP}) we employed an It\^o-Euler integrator~\cite{Gardiner2004}
with a time step $\Delta t$ ranging from $2\times 10^{-7}\cdot M^{-1}$ to $10^{-5}\cdot M^{-1}$ depending on the size of the
set $\mathcal{V}_B$ and on the number of switched off detectors. Such a choice enabled us to have an efficient integrator, even
though some of the quantum trajectories might have been unstable. To solve the instability problem, we have first tried to apply an implicit
Miltstien method~\cite{Gardiner2004,Kloeden1992}, but since both~(\ref{eq:sderk}) and~(\ref{eq:sdeP}) are nonlinear, one has to solve
numerically at each time step
(e.g., by means of the Nelder-Mead method~\cite{Press2007}) implicit equations like $\vec{r}_k(t+\Delta t)=\vec{r}_k(t)+f(\vec{r}_k(t+\Delta t))$,
where $f$ is the r.h.s. of Eq.~(\ref{eq:sderk}) plus some additional term due to the Miltstien routine. While such a strategy might solve
the problem, we have numerically observed that such an approach is significantly more  time consuming than the It\^o-Euler integrator.
Thus, we also applied a derivative free order 2.0 weak predictor corrector method~\cite{Kloeden1992}, which turns out to be quite efficient
in the case of a single Wiener noise process, but in the case of three detectors, whose generalization is not straightforward, we noticed, 
as for other predictor-corrector methods, that the instability could not be fixed. Hence, we employed the simple It\^o-Euler
integrator with (rather) small time steps (up to $\Delta t = 2\times 10^{-7}\cdot M^{-1}$). We noticed that with such a simple strategy the
unstable quantum trajectories could have been reduced or even suppressed, but at the expenses of a very long numerical computation.


\bibliography{Literatur_Magnetometry}

\end{document}